\documentclass[aps,prb,twocolumn,groupedaddress]{revtex4}

\newif\ifpdf
\ifx\pdfoutput\undefined
\pdffalse % we are not running PDFLaTeX
\else
\pdfoutput=1 % we are running PDFLaTeX
\pdftrue
\fi

\ifpdf
\usepackage[pdftex]{graphicx}
\else
\usepackage{graphicx}
\fi
\usepackage{hyperref}

\begin{document}

\ifpdf
\DeclareGraphicsExtensions{.pdf, .jpg}
\else
\DeclareGraphicsExtensions{.eps, .jpg}
\fi

\def\hslash{\hbar}
\def\imag{i}
\def\grad{\vec{\nabla}}
\def\div{\vec{\nabla}\cdot}
\def\curl{\vec{\nabla}\times}
\def\DDt{\frac{d}{dt}}
\def\ddt{\frac{\partial}{\partial t}}
\def\ddx{\frac{\partial}{\partial x}}
\def\ddy{\frac{\partial}{\partial y}}
\def\lap{\nabla^{2}}
\def\divv{\vec{\nabla}\cdot\vec{v}}
\def\gradS{\vec{\nabla}S}
\def\vvec{\vec{v}}
\def\wc{\omega_{c}}
\def\<{\langle}
\def\>{\rangle}
\def\Tr{{\rm Tr}}
\def\Csch{{\rm csch}}
\def\Coth{{\rm coth}}
\def\Tanh{{\rm tanh}}
\def\g2{g^{(2)}}

% Use the \preprint command to place your local institutional report
% number in the upper righthand corner of the title page in preprint mode.
% Multiple \preprint commands are allowed.
% Use the 'preprintnumbers' class option to override journal defaults
% to display numbers if necessary
%\preprint{}

%Title of paper
\title{Current producing states in molecular semiconductors: photo-current from a 
molecular wire}
% repeat the \author .. \affiliation  etc. as needed
% \email, \thanks, \homepage, \altaffiliation all apply to the current
% author. Explanatory text should go in the []'s, actual e-mail
% address or url should go in the {}'s for \email and \homepage.
% Please use the appropriate macro foreach each type of information

% \affiliation command applies to all authors since the last
% \affiliation command. The \affiliation command should follow the
% other information
% \affiliation can be followed by \email, \homepage, \thanks as well.
\author{Eric R. Bittner}
\affiliation{Department of Chemistry and Center for Materials Chemistry\\
University of Houston, Houston, TX 77204}
\author{Aijun Ye}
\affiliation{Department of Chemistry and Center for Materials Chemistry\\
University of Houston, Houston, TX 77204}
\affiliation{Chemistry of Novel Materials,
University of Mons-Hainaut,
Place du Parc, 20; B-7000 Mons, Belgium}
\author{Stoyan Karabunarliev}
\affiliation{Department of Chemistry and Center for Materials Chemistry\\
University of Houston, Houston, TX 77204}

%Collaboration name if desired (requires use of superscriptaddress
%option in \documentclass). \noaffiliation is required (may also be
%used with the \author command).
%\collaboration can be followed by \email, \homepage, \thanks as well.
%\collaboration{}
%\noaffiliation

\date{\today}

\begin{abstract}
We present a methodology for computing photocurrent production in molecular 
semiconducting molecules.  Our model combines a single-configuration interaction 
picture with the Schwinger-Keldysh non-equilibrium Greens function approach to 
compute the current response of a molecular semi-conducting wire following excitation.
We give detailed analysis of the essential excitonic, charge-transfer, and dipole states for poly-(phenylenevinylene) chains of length 32 and 48 repeat units under 
an electric field bias and use this to develop a reduced dimensional tunneling model 
which accounts for chain-length and field-dependent behavior. 
\end{abstract}

% insert suggested PACS numbers in braces on next line
% insert suggested keywords - APS authors don't need to do this
%\keywords{}

%\maketitle must follow title, authors, abstract, \pacs, and \keywords
\maketitle

% body of paper here - Use proper section commands
% References should be done using the \cite, \ref, and \label commands
\section{Introduction}
% Put \label in argument of \section for cross-referencing
%\section{\label{}}

Advances in nanotechnology have lead to the fabrication of the devices which length-scales 
smaller than the mean free path of an electron.~\cite{Datta}
Work in this direction as pushed the scale of an individual ``device'' to the 
molecular scale through reported measurements of electronic transport
through carbon nanotubes, self-assembled monolayers of conjugated polymers, and 
indivisual molecules. ~\cite{Joachim}  Moreover, there have been recent measurements of
electrically induced light emission from individual single-walled carbon nanotubes~\cite{Avouris},
as well as incredible progress towards the fabrication and synthesis of direct-bandgap
nanowires and super-lattices with novel optical-electronic properties. 

 As this technology 
continues to press towards the molecular level, energy quantization, phase coherences, 
and electron-phonon coupling play increasingly important roles in the properties of the
device.  The central theme and challenge in designing molecular electronic components
is the manipulation and control of charge flow through single molecules and molecular
assemblages.   Predicting and understanding electronic current flow through molecular
systems from an atomistic and first principles point of view presents a formidable 
theoretical challenge in that in requires the extension of standard quantum chemical 
methods that are well suited for bound state problems to solve non-stationary and 
many-body scattering in the continuum. The difficulty arises in  how to impose 
open boundary conditions and steady flow-boundary conditions in a computationally
feasible way.   

By in large, the theoretical description of charge transport in a molecular device 
can be cast in the language of quantum scattering theory in both time dependent 
and time-independent forms.  The physical picture is that given by Landauer~\cite{landauer,Datta}
in which current through a molecular wire is generated by 
 ballistic charge carriers
scattering through the molecule.  Hence, conductivity is related to the transition matrix
matrix, $T$. 
For the time-independent case, the formal treatment
involves the use of non-equilibrium Keldysh-Green function treatments
to compute the $T$-matrix.~\cite{Lang,Nitzan,NitzanRatner}   
Within the ``scattering'' approach, one derives a self-energy contribution to the molecular 
Hamiltonian due to the contact with the leads attached to the wire.   This leads to a non-Hermitian
Hamiltonian with complex eigenvalues.  
In contrast to the Landauer approach, Kosov\cite{kosov1,kosov2,kosov3} establishes a molecular wire as a non-equilibrium
steady-state system by appending a  Lagrange multiplier to the molecular Hamiltonian
to constrain the current passing through the system to be some desired value.  As a 
consequence, the 
modified Hamiltonian remains Hermitian, continuity is rigorously enforced, and 
the bias is determined as the difference between the Fermi energies of the system 
under forward and reverse  current. 

In this work, we focus upon electronic transport through semiconducting systems
where Coulomb interaction and electron correlation effects become important.
We address the issue of current due to photoexcitation of a conjugated polymer
``wire''.   We focus on the case where the bias applied to the semiconductor
is insufficient to produce current so that all of the charge-carriers in the system 
are due to the photo-excitation process.  We adopt the Newn-Anderson~\cite{Newns,Anderson}
model for a two band-semiconducting polymer wire and use this description to 
compute the current response of the wire.   

\section{Theoretical Model} 

\begin{figure}
\includegraphics[width=3.5in]{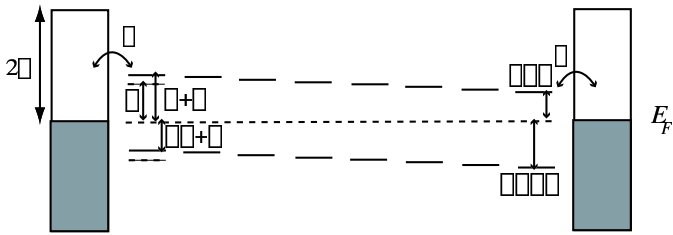}
\caption{Scheme of the one-particle description and definition of injection energy with respect
to the band-center.}\label{scheme}
\end{figure}

\subsection{Metal-Polymer-Metal Hamiltonian}
We consider here a rather idealized scenario in which an $N$-site polymer chain
bridging two metal electrodes.   This is idealized in the generic
sense because in the majority  of photo-current experiments, the polymer chromophores
in the sample are either randomly distributed in the sample or have more or less random 
orientation with respect to the applied bias field.  Moreover, with the exception of 
experiments on single molecules or molecules oriented on a surface, the molecules themselves
are not chemically bound to the electrodes.  Nonetheless, we recognize these as possible
limitations of our theoretical model.   The formalism we give here is adapted from the typical single-particle 
perspective for the case of electron/hole interactions.  By assuming that the coupling to the 
continuum is a single-particle interaction, we can adopt a general scattering-theory perspective 
to describe current producing state in the presence of electron/hole excitations.

To begin, we partition the state space of
our system into three domains, $Q_{L\alpha}$ and $Q_{R\alpha}$ which span the
 states of the left and right electrodes 
and $P$ which spans the electronic states of the bridging molecule
and be projected into separate Fock spaces representing different total numbers of
elementary electron/hole excitations. 
$$
P = P_0 + P_1 + P_2 + \cdots
$$
Thus, the full Hamiltonian has the structure
\begin{eqnarray}
H = \left(\begin{array}{ccc}
H_{L\alpha}  & V_{PL} & 0 \\
V_{PL}        & H_P & V_{PR} \\
0       &  V_{PR}     & H_{L\alpha} 
\end{array}
\right)
\end{eqnarray}
where the diagonal terms are the hamiltonians for the uncoupled subsystems
and $V_{PL}$ and $V_{PR}$ are the interactions between the polymer and the 
left and right electrodes.   We will assume that number of excitations within the 
total system is conserved so that $V_{PL}$ and $V_{PR}$ cannot couple 
different polymer Fock spaces and are thus single particle (electron or hole) operators.

Thus, within the space of single excitations, the polymer
Hamilton is given by\cite{karabuna1,karabuna2,karabuna3} 
\begin{eqnarray}
H_{P} = \sum_{nm}( F_{mn} + V_{mn})  |m \rangle\langle n| 
\end{eqnarray}
where $|m\rangle = |m_e \overline{m}_h\rangle$ is mono-excited electron/hole 
configuration, $V_{mn}$ is the two-particle matrix-element, 
and $F_{mn}$ is the matrix of the one-particle band-structure operator, $\hat{f}$
(which also includes any external field bias).   In the absence 
of the electrodes, this is given by 
\begin{eqnarray}
F_{mn} = \delta_{m_hn_h} \langle m_e | \hat{f} | n_e\rangle - \delta_{m_en_e}
 \langle m_h | \hat{f} | n_h\rangle
\end{eqnarray}
When the coupling to the electrodes is introduced, these terms need to be modified to 
include the one-particle polymer-electrode coupling.   For this we need to make a 
number of assumptions regarding the form of this coupling.  First, we assume that 
the polymer-electrode coupling occurs only at the terminal sites of the polymer.  
\begin{eqnarray}
V_{LP} = \eta_{km_em_h} 
( |k\rangle\langle 1_e\overline{m}_h| + |k\rangle\langle m_e\overline{1}_h|).
\end{eqnarray} 
Using the Feshbach method~\cite{Joachain,Datta,Datta2} 
and assuming that the density of states of the electrodes
is that of a 3D metal, we can derive reduced one-particle matrix elements, $\tilde{F}$
for the terminal sites of the polymer.
\begin{eqnarray}
\tilde{F}_{11} = F_{11} - \Sigma_1(E) \\
\tilde{F}_{NN} = F_{NN}-\Sigma_N(E)
\end{eqnarray}
with the remaining terms left unmodified.  $\Sigma^R_{K}(E) = \Delta_K(E) - i \Lambda_K(E)$ 
is the complex (retarded) self-energy contribution from the electrodes.     
(Note: $\Sigma^A_K = (\Sigma^R_K)^*$ is the advanced self-energy.)
For these, we adopt the Newns-Anderson~\cite{Newns,Anderson} model  
and write
\begin{eqnarray}
\Lambda_K(E) &=& \frac{\beta^2_K}{\gamma}\left\{
\begin{array}{ll}
\frac{E}{2\gamma},  &  \left|E/2\gamma\right| < 1 \\
\frac{E}{2\gamma}  + \sqrt{\left(\frac{E}{2\gamma}\right)^2-1}, & {E}/{2\gamma} < 1\\
\frac{E}{2\gamma}  - \sqrt{\left(\frac{E}{2\gamma}\right)^2-1} ,  & {E}/{2\gamma} > 1
\end{array}
\right. \\
\Delta_K(E) &=&
 \frac{\beta^2_K}{\gamma}\left\{
\begin{array}{ll}
\sqrt{1-\left(\frac{E}{2\gamma}\right)^2}   &  \left|E/2\gamma\right| < 1 \\
0, &\mbox{otherwise}.
\end{array}
\right.
\end{eqnarray}
where $E$ is the ``injection energy'' for an electron (or hole) to tunnel between  the metal 
and the polymer measured relative to a common Fermi energy, $E_F$,
 $4\gamma$ is the band-width of the
metal reservoir, and $\beta_K$ is the chemisorption coupling between the polymer and 
electrode.  This system is illustrated in Fig.~\ref{scheme}.   For a metal/polymer contact, 
we take $4\gamma = 40$ eV as  the band-width of the electrode and
with a chemisorption coupling of $\beta_K = 0.5$eV.   For the unbiased case, 
the injection energy for an electron or a hole is half the band-gap of the polymer, 
$E = 2.5 $eV for PPV so that both the bonding and anti-bonding orbitals of the 
$\pi$-electron system fall into the band-width of the metal.   Thus, $|E/2\gamma| < 1$
and we can approximate the retarded self energy as 
\begin{eqnarray}
\Sigma_K^R(E) \approx - i  \frac{\beta^2_K}{\gamma}
\end{eqnarray}
which indicates that the coupling to the electrode should produce only  a minor 
perturbation to the real energy of the system and that its only effect is to provide
a sink for the e/h excitation to decay. We will also assume throughout
that the applied bias is insufficient to cause direct current flow through the system. 
The Newns-Anderson model was also used by Mujica {\em et al.}~\cite{mujica1,mujica2}  for the description of scanning 
tunneling microscopy (STM) current in molecular imaging of one-dimensional 
systems to encompass the more general process of electron transfer between two
reservoirs. 

Our primary assumption throughout is that the self-energy contribution is a 
single-particle term.   Thus, we can include the effect of the coupling  to the metal
leads into a configuration interaction theory  by including this term in the 
generation of the Fock-matrix
\begin{eqnarray}
F_{{\bf mn}} = \delta_{\overline{m}\overline{n}} \langle m | (\hat{f} + \Sigma)| n\rangle - \delta_{mn}
 \langle \overline{m} |( \hat{f}+ \Sigma) | \overline{n}\rangle 
\end{eqnarray}
where 
\begin{eqnarray}
\langle m | \Sigma^R |n\rangle &=& \delta_{mn}(\delta_{n1}\Sigma^R_1(\varepsilon+\mu) +\delta_{nN}\Sigma^R_N(\varepsilon-\mu)\nonumber \\
\langle \overline{m} | \Sigma^R |\overline{n}\rangle &= &
\delta_{\overline{m}\overline{n}}(
\delta_{\overline{n}\overline{1}}\Sigma^R_1(-\varepsilon-\mu)+\delta_{\overline{n}\overline{N}}\Sigma^R_N(-\varepsilon-\mu) )\nonumber
\end{eqnarray}
were $\pm\varepsilon$ is the injection energy for an electron (+)  or hole (-) onto the terminal site  in the absence of an applied
bias and $\mu$ is the shift of the bare site energy due to the applied field.   For this we take the Fermi energy of the 
system as a common reference.  
Including the self-energy into mix produces a non-Hermitian complex symmetric 
CI Hamiltonian which can be diagonalized via unitary transformation yielding
complex eigenvalues, $\tilde\varepsilon_\alpha$,
\begin{eqnarray}
(H_{P}+\Sigma^R)|\varphi_\alpha\> = \tilde\varepsilon_\alpha|\varphi_\alpha\>.
\end{eqnarray}

Since the polymer/electrode interaction is at the terminal ends and only involves single-electron
terms, we have an exact relation between the net electron/hole population for a given CI state
 on the terminal sites and  complex energy shift component of the CI eigenvalue, $\tilde\varepsilon_\alpha
 = \varepsilon_\alpha + \delta\epsilon_\alpha + i \lambda_\alpha$ where the real energy shift 
and imaginary energy are related to the real and imaginary components of the 
self-energy, $\Sigma_K$. 
\begin{eqnarray}
\delta\epsilon_\alpha &=&  \Lambda_K
( e_{L\alpha}^- + e_{R\alpha}^- + h_{L\alpha}^+ + h_{R\alpha}^+ )\\
\lambda_\alpha &=&  \Delta_K
( e_{L\alpha}^- + e_{R\alpha}^- + h_{L\alpha}^+ + h_{R\alpha}^+ ) 
\end{eqnarray}
where, 
\begin{eqnarray}
e_{L\alpha}^- = \sum_{\overline{m}}|\langle 1\overline{m}|\varphi_\alpha\rangle|^2  \\
h_{L\alpha}^+ = \sum_{m}|\langle m\overline{1}|\varphi_\alpha\rangle|^2  \\
e_{R\alpha}^- = \sum_{\overline{m}}|\langle N\overline{m}|\varphi_\alpha\rangle|^2  \\
h_{R\alpha}^+ = \sum_{m}|\langle m\overline{N}|\varphi_\alpha\rangle|^2  
\end{eqnarray}
are the populations of the electron or hole on the terminal (i.e. right and left) ends of the polymer.
In essence, eigenstates with more total electron or hole population on the terminal sites
are more coupled to the metal continuum than eigenstates with less amplitude on the 
terminal sites.

\begin{figure}[b]
\includegraphics[width=3.5in]{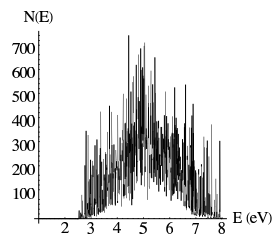}
\caption{Density of states for PPV$_{48}$ under 0.03eV/unit bias.}\label{DOS48}
\end{figure}

Once we have obtained the complex-eigenspectrum, we can construct the 
retarded Greens function for e/h scattering through the molecule.
\begin{eqnarray}
G^R(E) &=& (E-H+i\eta)^{-1} \nonumber \\ &=&(E-H_{P}-\Sigma^R)^{-1}.
\end{eqnarray}
Because $H_{P} + \Sigma^R$ is {\em not} a Hermitian operator, its 
eigenvectors do not form a orthogonal set and we need to compute the 
eigenspectrum of the adjoint matrix
\begin{eqnarray}
(H_{P}+ \Sigma^A)|\psi_\alpha\> = \tilde\varepsilon^*|\psi_\alpha\>.
\end{eqnarray}
Taken together, the eigenvectors $\{|\varphi_\alpha\>\}$ and $\{|\psi_\alpha\>\}$ form a bi-orthogonal
set
\begin{eqnarray}
\sum_{\bf n}\varphi_\alpha({\bf n})^*\psi_{\beta}({\bf n}) =  \delta_{\alpha\beta}.
\end{eqnarray}
Consequently, the retarded Greens function is constructed via
\begin{eqnarray}
G^R({\bf n},{\bf n'}) &=& \sum_\alpha \frac{\<{\bf n}|\psi_\alpha\>\<\varphi_\alpha|{\bf n'}\>}{E - \tilde\varepsilon_\alpha}.
\end{eqnarray}
In a similar vein, the spectral response, which gives the generalized density of states 
inside the polymer taking the contact with the metallic continuum into account is given by 
\begin{eqnarray}
A &=& i (G^R-G^A)  \nonumber \\
&=& G^R \Gamma G^A
\end{eqnarray}
where $\Gamma$ is the energy broadening taken as as the difference between the
 retarded and advanced self-energies
 \begin{eqnarray}
 \Gamma = i (\Sigma^R - \Sigma^A)
 \end{eqnarray}
 Pulling these two together yields the usual expression
 \begin{eqnarray}
 A({\bf n},{\bf n}') = \sum_a \lambda_a \frac{\varphi_\alpha({\bf n})\psi_\alpha({\bf n'})}{(E-\varepsilon_\alpha + \delta\epsilon_\alpha)^2+(\lambda_\alpha/2)^2}
 \end{eqnarray}
 where $\varepsilon_\alpha$  is the ``bare'' CI eigenvalue of $H_P$ in the absence of the metal contact.  Lastly, the 
 density of states is given by the trace of $A$,
 \begin{eqnarray}
 N(E) = \frac{1}{2\pi}\sum_a \frac{\lambda_a }{(E-\varepsilon_\alpha + \delta\epsilon_\alpha)^2+(\lambda_\alpha/2)^2}, 
\end{eqnarray}
Thus, each CI state in becomes broadened by the contact with the metal implying that the states take on a finite life-time 
in addition to being displaced in energy. 
In the case of weak interaction, $\lambda_\alpha \rightarrow 0$ and the ``Lorentzian'' describing the spectral contribution from the 
$\alpha$ state collapses to a $\delta$-function.
The density of states for a 32-repeat unit model of PPV under 0.03eV/unit bias 
is shown in Fig.~\ref{DOS48}

\begin{figure}[t]
\includegraphics[width=3.5in]{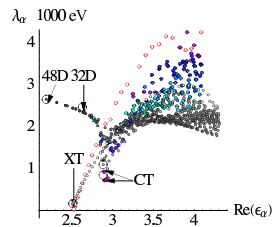}
\caption{(color) Complex energy distribution for the SCI states of
$PPV_{32}$ (colored disks) and $PPV_{48}$ (gray-disks) 
under 0.03eV/unit bias.
 Note the 
imaginary energy component $\lambda$ is scaled by $10^3$. 
The shading or hue for each point indicates the degree of polarization of each state as measured by the net charge on the terminal end of the chain with open circles indicating
no net charge on the terminal site (red for $PPV_{32}$). 
 The six circled energies correspond to the 
lowest energy dipole state (D), exciton (XT), and charge transfer (CT)
 state for either chain.
}\label{energies}
\end{figure}

In Fig.~\ref{energies} we compare
 the distribution of the real and imaginary components of the singlet SCI states 
for a PPV polymer chain of 48  and 32 repeat units with a bias potential of 
0.03eV/unit cell.  This corresponds to a macroscopic field of $1.0\times 10^5V/cm$
which is on-par with the fields placed across actual devices.  
%From this we see that lowest energy states have the smallest imaginary
%energy components.  In fact, the lowest energy state, corresponding to the 
%$1B_u$ exciton has almost no imaginary energy component what so ever.  
%If we plot the electron/hole density of this state we see that in fact the state is 
%localized entirely within the interior of the polymer chain and has negligible 
%density on the terminal sites.   
The  shading  of each point in the figure reflects the net charge on the 
terminal ends of the molecule with white or red circles indicating no  
net charge on the terminal site.  Also indicated on this plot are 
a set of energies we will associate with the lowest
 exciton (XT), the lowest energy pure charge transfer state where the 
 electron and hole do not reside on the same repeat unit (CT) , and a ``dipole''
 state (D) in which the electron and hole are localized on opposite ends of the chain. 

\begin{figure}[t]
\includegraphics[width=3.5in]{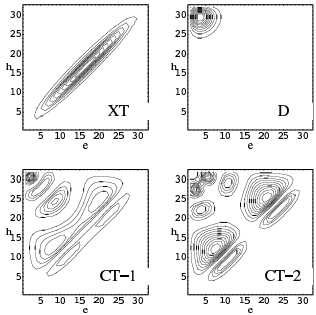}
\caption{ Electron/hole distributions for  the XT, D, and two CT states for 
$PPV_{32}$ at 0.03eV/unit bias.  The energies for these states are circled on 
Fig.~\ref{energies}}\label{distrib}
\end{figure}

We  see that the energies 
of the lowest energy excitonic state  are identical for 
both chains at 2.5eV as is the location of lowest pure charge-transfer state at 
about 2.9eV.  Both of these sets of states are indicated by circles on Fig.~\ref{energies}.
 What is interesting in this regard is that there appears to be a 
progression of purely excitonic states intersecting a progression of 
charge-separated states.   This progression of excitonic states are simply the
center-of-mass translational eigenstates of the lowest energy bound electron/hole 
state. 
Even though the exciton states acquire 
quite large (in comparison) imaginary energy components  indicating that the 
life-time of the states steadily decrease as energy increases, they remain 
relatively unpolarized by the applied field.  As we shall demonstrate 
next, the current produced by a given excitation is determined by both the 
size of the imaginary energy components and by the net charge on the 
terminal ends of the molecule.  Consequently, the excitonic states will contribute
very little to the current where as the charge-transfer states
will give the dominant contribution.  

\begin{figure}
\includegraphics[width=3.5in]{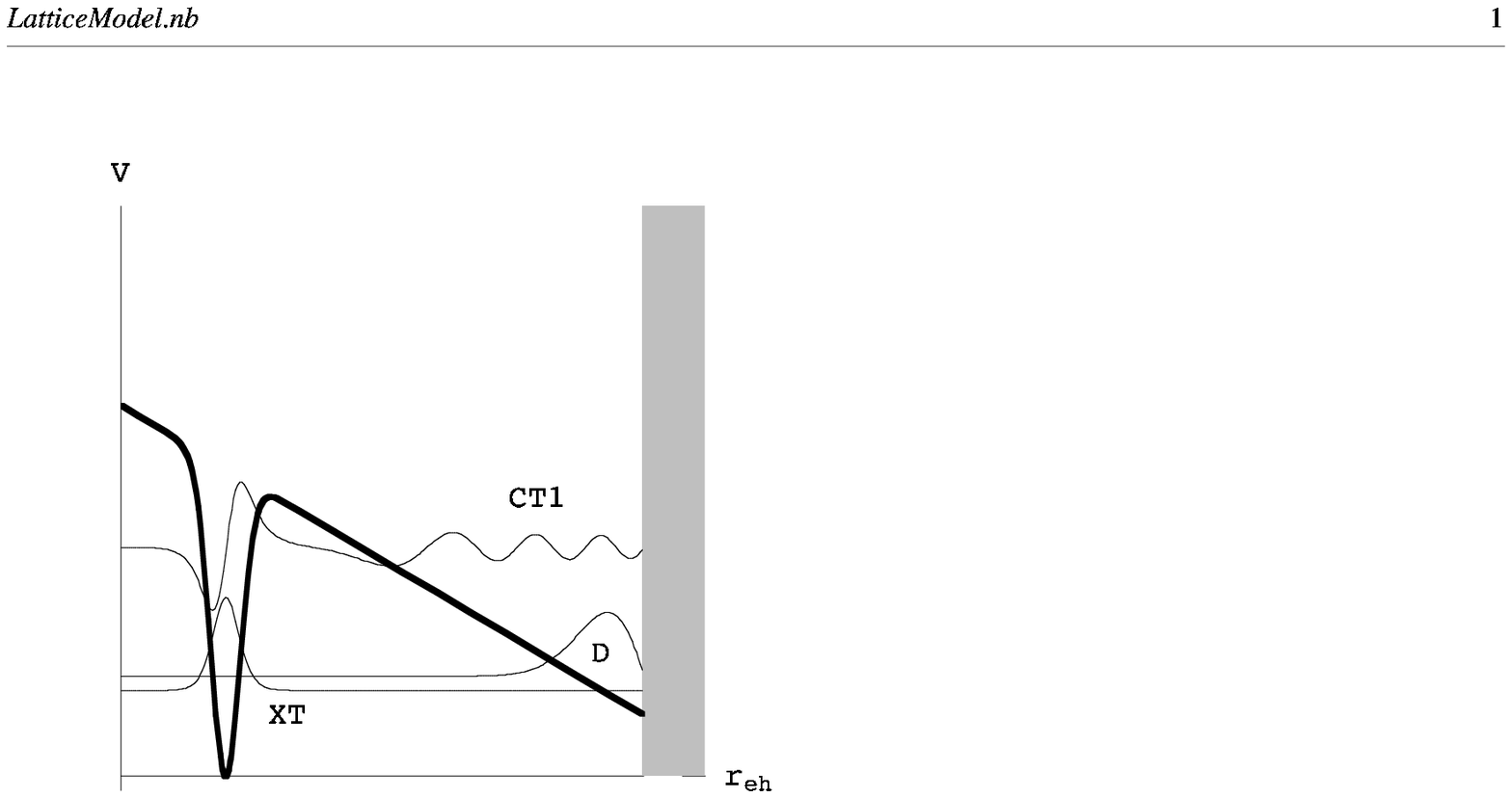}
\caption{One dimensional reduced model for unimolecular decay of a bound electron/hole pair in the presence of an electric field.  The total electron/hole potential 
is indicated by the thick line and the corresponding exciton, dipole, and charge-transfer
states are superimposed.  The shaded ``wall'' corresponds to the end of the polymer 
connected to the continuum.   As the length of the polymer increases, the $D$ state 
moves farther to the right and lower in energy whereas the XT and CT remain 
more or less unchanged.  }\label{model}
\end{figure}

In Fig.~\ref{distrib} we plot the XT, D, and first two CT states for the 
32 unit chain.  Here, the exciton  is more or less unperturbed by the 
applied field and is more or less identical to the exciton for the polymer in the
absence of the field.  For the dipole state (D), the electron and hole are localized 
on the opposite ends of the polymer.  For the 32 unit chain, $E(D) > E(XT)$; however, 
for the 48-unit chain the dipole state is considerably lower in energy than the 
XT.  The energy of this state should scale almost linearly with chain size 
since to lowest order approximation $E(D) = q n a$ as per a dipole in an electric 
field.  The CT states shown in Fig.~\ref{distrib} are  tunneling states from a bound CT state (i.e. the $mAg$) into the continuum.  In fact this entire picture can be 
rationalized in a simple one-dimensional model of the unimolecular decay of a 
bound electron/hole state as illustrated in Fig.~\ref{model}.

\subsection{Current response to an excitation}

If we consider the net flow of charge from the molecule following promotion to some state $\phi_\alpha$, 
we need to be cognizant of the fact that even though the lifetime of the state is given by $2\lambda_\alpha/\hbar$, this
represents only the net decay of the electron/hole excitation from the molecule into the metal 
since there is an equal influx and efflux of charge to and from the molecule.   
In other words, the current produced by a given state is not simply proportional to the 
rate of decay of the state  into the continuum.  Furthermore, in the limit that the valance and conduction 
bands have identical interaction with the metal (as in the case at hand) no net charging 
of the molecule can occur as a result of contact with the metal.  Hence, total charge $Q$ on the 
molecule must remain a constant of the motion.  Since $I = \dot Q$, one is tempted to 
conclude $I = 0$.  However, the net current should be taken as the difference between its
retarded and advanced components.
\begin{eqnarray}
I &=& (\dot Q^R - \dot Q^A) = i[\Sigma^R-\Sigma^A,Q]/\hbar\\
&=&e [\Gamma,Q]/\hbar
\end{eqnarray}
Thus, the current produced at a given excitation energy $E$ is given by 
\begin{eqnarray}
I(E) &=& {\rm Tr}( [\Gamma , Q])/\hbar \nonumber  \\
&=& \frac{e}{\hbar}\frac{1}{2\pi}
\sum_a 
\frac{\lambda_a^2 ((e_{La}^- - h_{La}^+)+ (h_{Ra}^+ -e_{Ra}^-))}{(E-\varepsilon_a+ \delta\epsilon_a)^2+(\lambda_a/2)^2} \label{dos-current}
\end{eqnarray}

For the case of electron/hole symmetry, we have the additional requirement 
\begin{eqnarray}
e^-_{L\alpha}-h^+_{L\alpha} =h_{R\alpha}^+-e^-_{R\alpha}.
\end{eqnarray} 
Thus, we can consider the current produced by a given state by writing
\begin{eqnarray}
I_\alpha 
&=& (2e\lambda_\alpha/\hbar)
\left(
\left(
\begin{array}{l}
\mbox{\tiny $e^-$ on site 1}\\
\mbox{\tiny hole anywhere}
\end{array}
\right)
-
\left(
\begin{array}{l}
\mbox{\tiny $h^+$ on site 1}\\
\mbox{\tiny $e^-$ anywhere}
\end{array}
\right)\right)\nonumber \\
&=& (2e\lambda_\alpha/\hbar)(e_{L\alpha}^- -h_{L\alpha}^+)
\end{eqnarray}
This is similar to the notion by Halpern that current carrying states are 
localized at the opposite ends of the conducting sample~\cite{Halpern}
since these are the states with the most interaction with the terminal leads. 
For purely excitonic states,  the net charge on the terminal sites will be exactly zero.  
Hence a purely excitonic species will produce no net current.   However, a 
charge-transfer species will produce current when the system is placed under a bias.
Reversing the bias produces current in the opposite direction 
since the current producing charge-transfer states will be polarized in the 
reverse direction.

\begin{figure}
\includegraphics[width=3.5in]{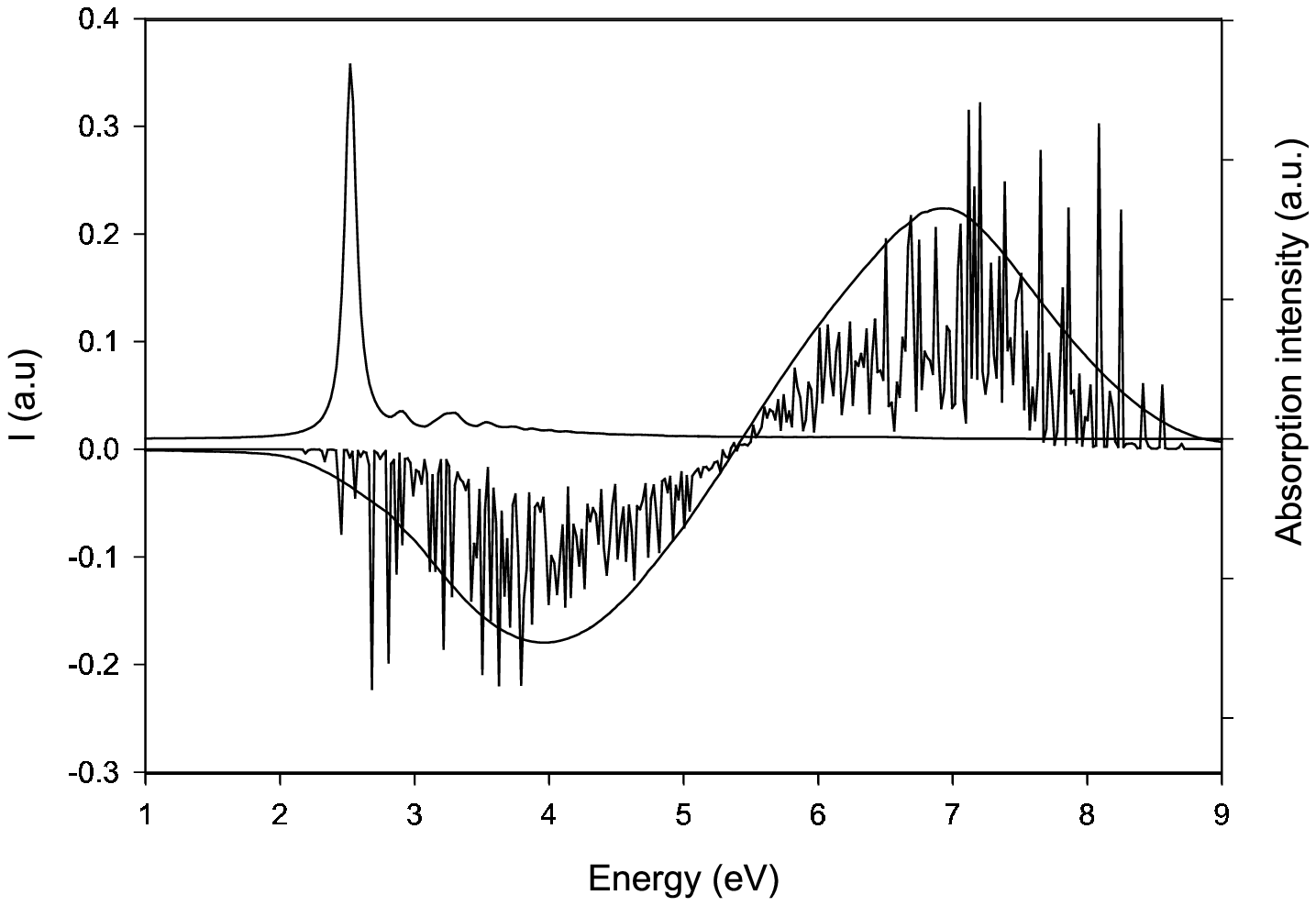}
\caption{Current vs. excitation energy and 
spectral response for $PPV_{48}$ at 0.03eV/unit bias.  
Here we show both the natural line-widths of the current resulting from a 
single chain and the inhomogeneously broadened current response. 
In this case, the effect of the field is to shift
 the lowest current producing state below the lowest energy exciton so that
 current is produced without any excess photon energy above the state 
 optically coupled to the ground state. }\label{current48}
\end{figure}

In Fig.\ref{current48} we show the current response for a given excitation energy
as well as the optical response for $PPV_{48}$ under 0.03eV/unit bias.  
From the way we have defined the direction of the current, a negative 
response indicates current in the forward direction, i.e. electrons flowing 
from the $-$ to $+$ ends and holes flowing $+$ to $-$.   
A zero response indicates that the 
electron current exactly balances the hole current, so no net charge flows 
through the system as a result of excitation.  At roughly 5.5 eV (center of the 
single-excitation band) the current response passes through zero and we see 
current going against the direction of the bias field as the energy increases.  Since 
this is well above the optical response of the system which peaks at about 2.5 eV, 
this reverse current most likely would not be observed in any molecular
system.  

%In this rather extreme case, the applied bias is sufficient to push the lowest energy current producing state below the optical absorption peak indicating that 
%a photo-excitation would produce current with out any excess energy. 
%However, the net current produced is quite small compared to currents produced 
%at energies above the peak in the optical response.  

\subsection{Effect of Bias Field}
\begin{figure}
\includegraphics[width=3.5in]{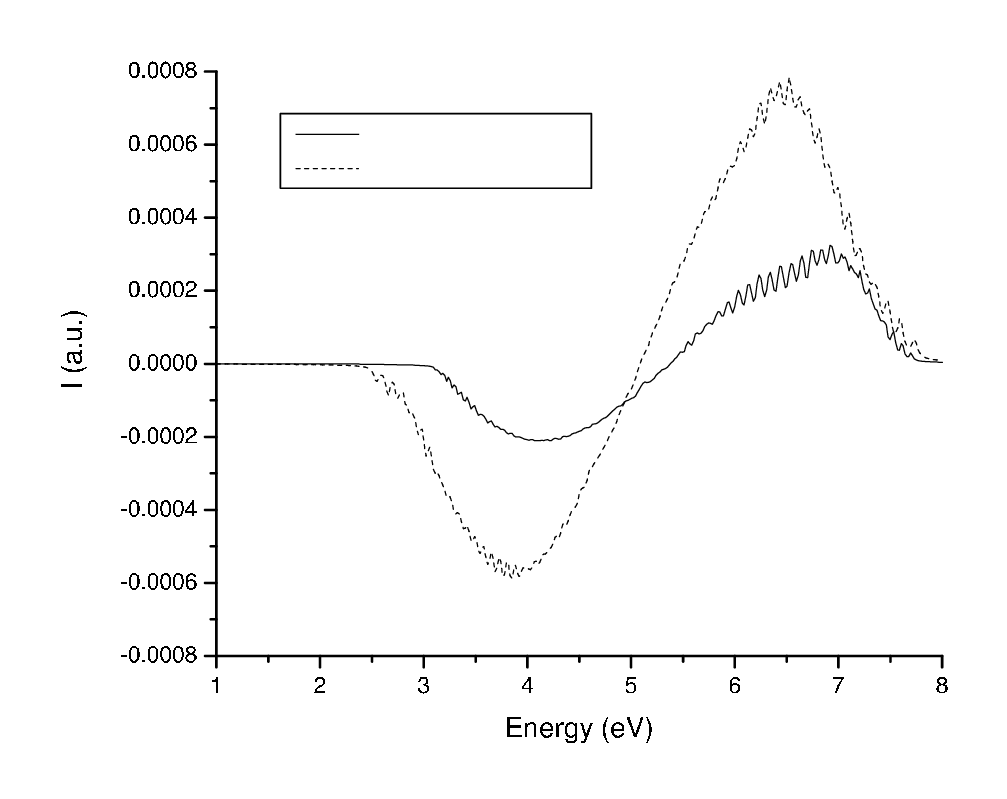}
\caption{Current vs. excitation energy for $PPV_{32}$ in two different applied fields (0.01eV/unit and 0.03 eV/unit).
 Here we have included a 100$\times\lambda_\alpha$ inhomogeneous line broadening  to each state to 
mimic the effects of environmental disorder.}\label{field}
\end{figure}

\begin{figure}
\includegraphics[width=3.5in]{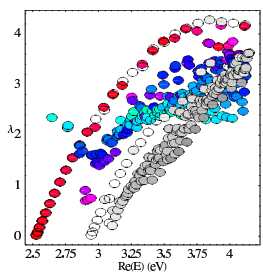}
\caption{Real and imaginary SCI energies for  $PPV_{32}$ in two different applied fields. Gray and open circles: 0.01eV/unit.  Red and hue-shaded disks: 0.03 eV/unit). }\label{field2}
\end{figure}

Finally we turn our attention towards what happens to the induced current as a 
function of applied field.  In Fig.~\ref{field} we show the current response
for a 32-unit chain under 0.01 eV/unit and 0.03eV/unit bias and the corresponding 
complex CI energies in Fig.~\ref{field2}. First we note that the
on-set of current shifts towards higher energies upon going from 0.03eV/unit to 0.01 eV/unit. 
This is commensurate with the
fact that as the field decreases, there is proportionally less Stark shifting of the 
dipole states.  
Consequently, at low field we have fewer potential current producing states at lower energies compared to the higher field case. 
On the other hand, the XT states remain completely unperturbed by the change
in field.  At lower field, the lowest CT states have smaller imaginary energies than 
at high-field.  Again, this is consistent with our reduced one dimensional model in 
Fig.~\ref{model} : lower-fields produce less tunneling between the CT states and the 
continuum.

\section{Discussion}

In this paper we have laid a theoretical ground-work for the study 
of current producing states in molecular semiconductors focusing 
solely up current produced upon electronic excitation.  In a subsequent
works, we will address the issue of DC and AC
conductivity in molecular semiconductors, especially those in which 
half of the molecule is more electron rich than the other. 
Such model pn-junctions have been produced synthetically
\cite{lu1,lu2} and we have
recently studied the electro-luminescence of such systems.\cite{karabuna4}

Needless to say, there are 
several lacuna in our  treatment that will need to be systematically addressed. 
First, we assume that the electric field across the molecule is uniform.  
In realistic systems, this may not be the case and one really should self-consistently
solve the Poisson equation for the field taking into account the actual charge
distribution of the molecule and the appropriate boundary conditions imposed by 
the conducting leads.   Similarly, the model fails to account for the image
charges between the polymer and metal contact.  Both of these effects 
require more detailed dynamical interactions between the metal leads and the polymer 
than is accounted for in the present model.   Secondly, we do not currently
account for lattice excitations or distortions due to the applied field or the 
current through the system. 
 The current response as we have defined it, simply gives  the 
coherent ballistic current produced at some energy $E$ above the HF ground state. 
We are currently working to extend this to finite temperature by the 
inclusion of phonon creation and annihilation channels. 

In spite of these shortcomings, the basic physical picture offered by this
model is quite appealing.  First of all, it emphasizes the fact that different electron/hole
states have widely differing capacities for producing current following a given 
excitation.  Possessing a large imaginary energy alone is insufficient to produce current. 
The state must also be sufficiently polarized so that there is adequate charge 
separation.  This raises another crucial aspect in trying to use this model 
to understand photocurrent in bulk systems.   That is, states with large dipole moments
will be poorly coupled to the ground state via optical excitation.  Consequently, 
direct excitation from the ground state to a strongly current producing state is
highly unlikely.  What is more likely is that following photo-excitation to some 
higher-lying eigenstate, phonon creation kicks in and induces transitions between the 
initial state and lower energy electronic states including those capable of 
producing current.

% Surround figure environment with turnpage environment for landscape
% figure
% \begin{turnpage}
% \begin{figure}
% \includegraphics{}%
% \caption{\label{}}
% \end{figure}
% \end{turnpage}

\begin{acknowledgments}
Funds for this work were provided by the National Science Foundation  (CAREER Award) and by the 
Robert A. Welch Foundation. 
\end{acknowledgments}


\begin{thebibliography}{10}

\bibitem{Joachim}
C. Joachim, J. K. Gimzewski and A. Aviram, Nature {\bf 408}, 541 (2001).

\bibitem{Datta}
{\em Electronic transport in mesoscopic systems}, S. Datta (Cambridge Univ. Press, New York, 1995).



\bibitem{Avouris}
J. A. Misewich, R.  Martel, Ph. Avouris, J. C. Tsang, S. Heinze, and J. Tersoff, 
Science {\bf 300}, 783 (2003).

\bibitem{Nitzan}
A. Nitzan, Annual Reviews of Phys. Chem. {\bf 52}, 681 (2001).

\bibitem{landauer}
R. Landauer, Philos. Mag. {\bf 21}, 863 (1970).

\bibitem{Lang}
N. D. Lang, Phys. Rev. B {\bf 52}, 5335 (1995).

\bibitem{NitzanRatner}
A. Nitzan and M. A. Ratner, Science {\bf 300}, 1384 (2003).

\bibitem{Ratner} 
Y. Xue, S. Datta, M. A. Ratner, J. Chem. Phys. {\bf 115}, 4295 (2001).



\bibitem{kosov1}
 D. S. Kosov, J. Chem. Phys. {\bf 116}, 6368 (2002).
 
 \bibitem{kosov2}
 D. S. Kosov and J. C. Greer, Phys. Lett. A {\bf 291}, 46 (2001).
 
 \bibitem{kosov3}
 D. S. Kosov, J. Chem. Phys. {\bf 119}, 1 (2003).


\bibitem{Joachain}
{\em Quantum Collision Theory}, C. J. Joachain (North-Holland, Amsterdam, 1975).


\bibitem{mujica1}
V. Mujica, M. Kemp, and M. Ratner, J. Chem. Phys. {\bf 101}, 6849 (1994). 

\bibitem{mujica2}
V. Mujica, M. Kemp, A. Roitberg and M. Ratner, J. Chem. Phys. {\bf 104} , 7296 (1996). 

\bibitem{Datta2} Y. Xue, S. Datta and M. A. Ratner, 
%First-principles based matrix
%Green's function approach to molecular electronic devices: general
%formalism, 
Chemical Physics {\bf 281}, 151 (2002).


\bibitem{heeger} 
A. J. Heeger in {\em Primary photo excitations in conjugated polymers}, N. S. Sariciftci, Ed. (World Scientific, Singapore 1997).

\bibitem{karabuna1}
S. Karabunarliev and E. R. Bittner, J. Chem. Phys.{\bf 118}, 4291 (2003). 

\bibitem{karabuna2}
S. Karabunarliev and E. R. Bittner, J. Chem. Phys. {\bf 119}, 3988 (2003). 

\bibitem{karabuna3} 
S. Karabunarliev and  E. R. Bittner, Phys. Rev. Lett.  {\bf 90} ,  057402 (2003).

\bibitem{Newns}
D. M. Newns, Phys. Rev. {\bf 178} , 1123 (1969). 

\bibitem{Anderson}
P. W. Anderson, Phys. Rev. 124, 41 (1961). 

\bibitem{may}
V. May and O. K\"{u}hn, {\em Charge and Energy Transfer Dynamics in Molecular Systems}, (Wiley -VCH, Berlin, 2000).

\bibitem{Halpern}
B. I. Halpern Phys. Rev. B {\bf 38}, 2185 (1982).

\bibitem{lu1}
  M.-K. Ng and L. Yu, 
 %Synthesis of amphiphilic conjugated diblock oligomers as molecular diodes, 
 Angewandte Chemie, International Edition
{\bf 41}, 3598 (2002).

\bibitem{lu2}
 M. K. Ng, D. C. Lee and L. Yu, 
 %Molecular Diodes Based on Conjugated Diblock Co-oligomers,
  J.  Am. Chem. Soc. {\bf 124},
11862 (2002).

\bibitem{karabuna4}
S. Karabunarliev and E. R. Bittner, J. Phys. Chem. A (submitted-2003)


%\bibitem{ref77} 
%B. R. Bulka and S. Lipinski,
%% Electronic Correlations in
%%Spin-Dependent Transport Through Nanostructures, 
%J. Superconductivity {\bf 16}, 275 (2003).

%\bibitem{ref78}
%P. Stefanski and B. R. Bulka, 
%%Electronic transport through large
%%quantum dots in the Kondo regime, 
%Physica Status Solidi B: Basic Research {\bf  236}, 388 (2003).

%\bibitem{ref79} 
%B. R. Bulka and S. Lipinski, 
%%Coherent electronic transport and
%%Kondo resonance in magnetic nanostructures,
% Phys. Rev. B {\bf 67}, 024404/1 (2003).


\end{thebibliography}
\end{document}